\documentclass[aip,jcp,showpacs,10pt]{revtex4-1}  

\usepackage[pdftex]{graphicx}
\usepackage{multirow}

\begin{document}


\title{Strong Correlations in Actinide Redox Reactions}


\author{S. E. Horowitz}
\email[]{witz@brown.edu}
\author{J. B. Marston}
\email[]{marston@brown.edu}
\homepage[]{http://www.brown.edu/Research/Environmental_Physics}
\affiliation{Department of Physics, Brown University, Providence, Rhode Island, 02912-1843 USA}



\begin{abstract}
Reduction-oxidation (redox) reactions of the redox couples
An(VI)/An(V), An(V)/An(IV), and An(IV)/An(III),  
where An is an element in the family of early
actinides (U, Np, and Pu), as well as
Am(VI)/Am(V) and Am(V)/Am(III), are modeled by combining density
functional theory with a generalized Anderson impurity 
model that accounts for the strong correlations between the 5f electrons.
Diagonalization of the Anderson impurity model yields improved estimates for the redox
potentials and the propensity of the actinide complexes to disproportionate.
\end{abstract}

\pacs{31.10.+z,31.15.aq,31.15.E-,31.70.Dk}

\maketitle


\section{Introduction}
\label{sec:introduction}
Chemical reactions of the early actinide elements in aqueous solution
are complex and challenging to predict.  Elements U, Np, and Pu 
each have four or more oxidation states in acidic
environments (Am has three).  Two of these states (III and IV) are
hydrated An$^{3+}$  and An$^{4+}$ ions; the other two (V and VI) form
linear \emph{trans}-dioxo AnO$_2^+$ and AnO$_2^{2+}$ actinyl
complexes. Disproportionation reactions are 
common, especially in the case of plutonium for which three of the
redox potentials are nearly the same (about one volt).  As different
oxidation states have widely different solubilities\cite{Fanghanel:2002},
redox chemistry plays a crucial role in the environmental dispersal of
actinides.\cite{Choppin:2001}  The rich behavior of actinide ions may be traced to the
valence electrons, especially those in the 5f shell.\cite{Clark:2000}
Because the 5f orbitals are relatively localized, the Coulomb
interaction induces strong correlations between electrons in the
shell.  The possibility that strong correlations in these solvated actinide and actinyl ions
enhance tendencies to disproportionate is an intriguing
hypothesis.  Support for this idea is provided by a Hubbard model of
5f electrons that disproportionates when solved in the Hartree-Fock (HF) approximation.\cite{Runge:2004p334}   

Quantitative models of actinide reactions must overcome several
obstacles.\cite{Schreckenbach:2010p126}  First relativity and the energetics of
solvation must be taken into account.  At present only the density functional theory (DFT)
method is capable of modeling these aspects accurately.   However, 
incorporating the physics of strong electronic correlations among the
5f electrons presents a greater challenge as these are known to be
poorly captured by DFT.\cite{Roberto:2006p44}
In this paper we take a hybrid approach to solving these problems by
using DFT to construct a generalized
Anderson impurity model of the frontier orbitals.\cite{Straus:1995p464,Hubsch:2006,Labute:2002,Labute:2004,Marston:1993p450,Onufriev:1996p451}  
Exact diagonalization of the Anderson impurity model corrects
the free energy obtained from DFT alone, yielding improved predictions
for redox free energies.  We emphasize that the hybrid method outlined here differs markedly from
the LDA+U approach as it does not simply modify the LDA functional to partly account 
for the Coulomb repulsion; rather a high-dimensional many-electron Hamiltonian that models
the physics of strong correlations between the 
important low-energy states is diagonalized.  The  approach also differs 
from configuration-interaction (CI) and its variants\cite{Bartlett:2007p315} in two significant ways:
First we are able to exactly diagonalize the low-energy effective model
with no restrictions placed on the ground state wavefunction.  
Second, the two-body Coulomb interaction is not the bare
electron-electron repulsion but rather an effective interaction that 
takes into account the effects of screening. 

The rest of the paper proceeds as follows:
In Sec. \ref{sec:DFT} we discuss the ab initio part of the calculation and
compare the results we obtain to calculations by other workers.
The independent-particle model that describes the Kohn-Sham (KS) orbitals
and the spin-orbit interaction is introduced in
Sec. \ref{sec:independentParticle}.  Incorporation of the effective Coulomb
interaction via a many-body model of the low-energy degrees of freedom is carried out in
Sec. \ref{sec:manyBody}.  The physics of strong
correlations are illustrated with the use of a simplified 
model in Sec. \ref{sec:fractionalOccupancy}.  The calculation of redox potentials and
other observables by the hybrid approach are presented in
Sec. \ref{sec:results}.  We conclude with some discussion in
Sec. \ref{sec:conclusion}.

\section{Density Functional Theory}
\label{sec:DFT}
DFT based studies of early actinides in aqueous solution have 
modeled the structure, vibrational frequencies, and free energies of
hydration.\cite{Schreckenbach:1998,Spencer:1999,Blaudeau:1999p463,Hay:2000,Tsushima:2001,Vallet:2001a,
Vallet:2004p370,Tsushima:2005,Sonnenberg:2005,Cao:2005,Shamov:2005,Gutowski:2006}
We do not attempt to review this work comprehensively here but refer the reader instead to 
Ref. \onlinecite{Schreckenbach:2010p126} and references therein.  As there is
significant hybridization between the higher orbitals, accurate
calculations require full quantum mechanical treatment of all orbitals
with principal quantum number $n = 5$ and higher.\cite{Kuchle:1994,Odoh:2010p124}  
Using small cores, ab initio calculations performed by Shamov and Schreckenbach\cite{Shamov:2005} were able
to reproduce An(VI)/An(V) redox potentials to within $0.6$ volt of
the measured values.   

The Amsterdam Density Functional (ADF)\cite{ADF} is an attractive DFT
package for modeling  actinides in solution because it includes
relativistic corrections via the zeroth-order regular approximation
(ZORA)\cite{van-Lenthe:1993,van-Lenthe:1996b}, uses a basis of
localized Slater-type atomic orbitals, and models solvation with
the Conductor like Screening Model (COSMO).\cite{Klamt:1995}  In our calculations the
first coordination sphere of water molecules are treated quantum
mechanically; COSMO is used to simulate a bulk dielectric medium
beyond the sphere.  Quantum mechanical modeling of the second
sphere of hydration may lead to improved agreement with
experiment\cite{Gutowski:2006} but we defer that for future work.  For
the sake of simplicity the number of water molecules in the first
solvation sphere are kept constant for each oxidation state: Eight each in the
case of An(III) and (IV) and five for AnO$_2$(V) and (VI).  (As discussed later,
we find that coordinating Pu(III) with 9 water molecules makes only a small difference.)
We use the revised PBE exchange-correlation
functional.\cite{Hammer:1999,Perdew:1996,Vosko:1980}   For the
actinides we employ a basis set modified from ADF's  triple-$\zeta$, 
doubly polarized (TZ2P) ZORA wavefunctions.  The frozen core consists of the 60 orbitals
for which $n \leq 4$.  The ADF-supplied basis 
set has 78 frozen core orbitals; 60 of these are retained in the core and the 5s, 5p, and 5d
orbitals are promoted to valence orbitals.  The number of sets of fit functions is accordingly 
increased from 79 to 87.
For oxygen and hydrogen the relativistic
TZ2P all-electron basis set provided by ADF is employed with no modification.

Spin-unrestricted calculations are carried out in three stages.  As a
first step the geometry of the actinide (or actinyl) plus the first
solvation sphere is optimized in the gas-phase (no COSMO).  We verify
that all vibrational modes have real frequencies at the optimal
geometry; these frequencies are used later in the thermodynamic
calculations.   The geometry is then allowed to relax in solvation
within the COSMO approximation for the surrounding dielectric, with
the following cavity radii: 1.350 \AA~ (H), 1.517 \AA~ (O) and 2.10 \AA~
(An).  Thermodynamic properties are calculated in the ideal gas
approximation with an effective pressure of 1354 atmospheres to
account for the reduced entropy of translation as appropriate for the
aqueous environment.\cite{Martin:1998}   As a final step an
independent-particle model of the frontier orbitals is constructed
(see Sec. \ref{sec:independentParticle}) and the single-electron
spin-orbit interaction is added to correct the electronic
contribution to the free energy.  

\begin{figure}
\centering
\includegraphics[width=120mm]{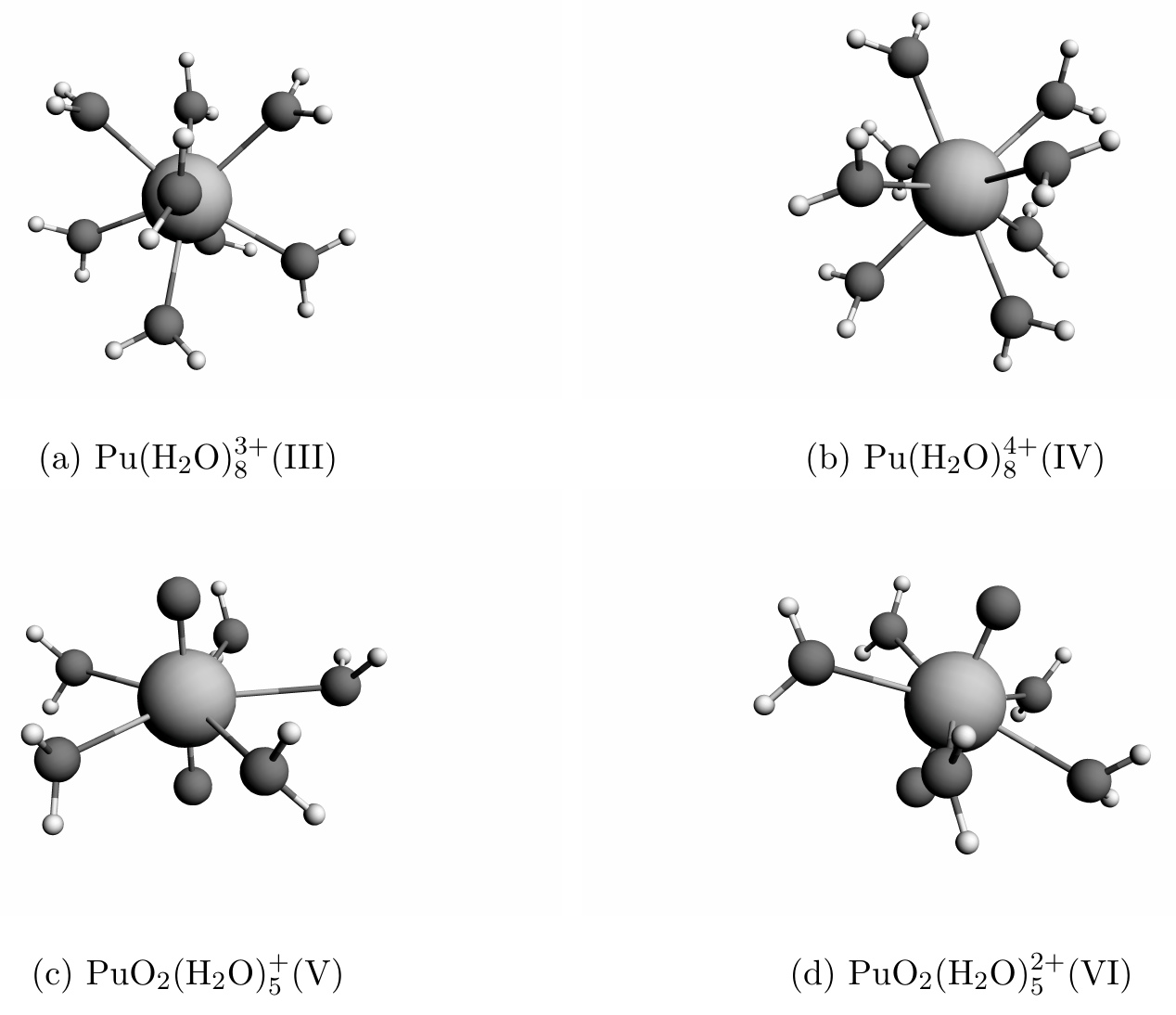} 
\caption{Converged geometries of the plutonium complexes.}
\label{fig:puGeo}
\end{figure}

\begin{figure}
\centering
\includegraphics[width=80mm]{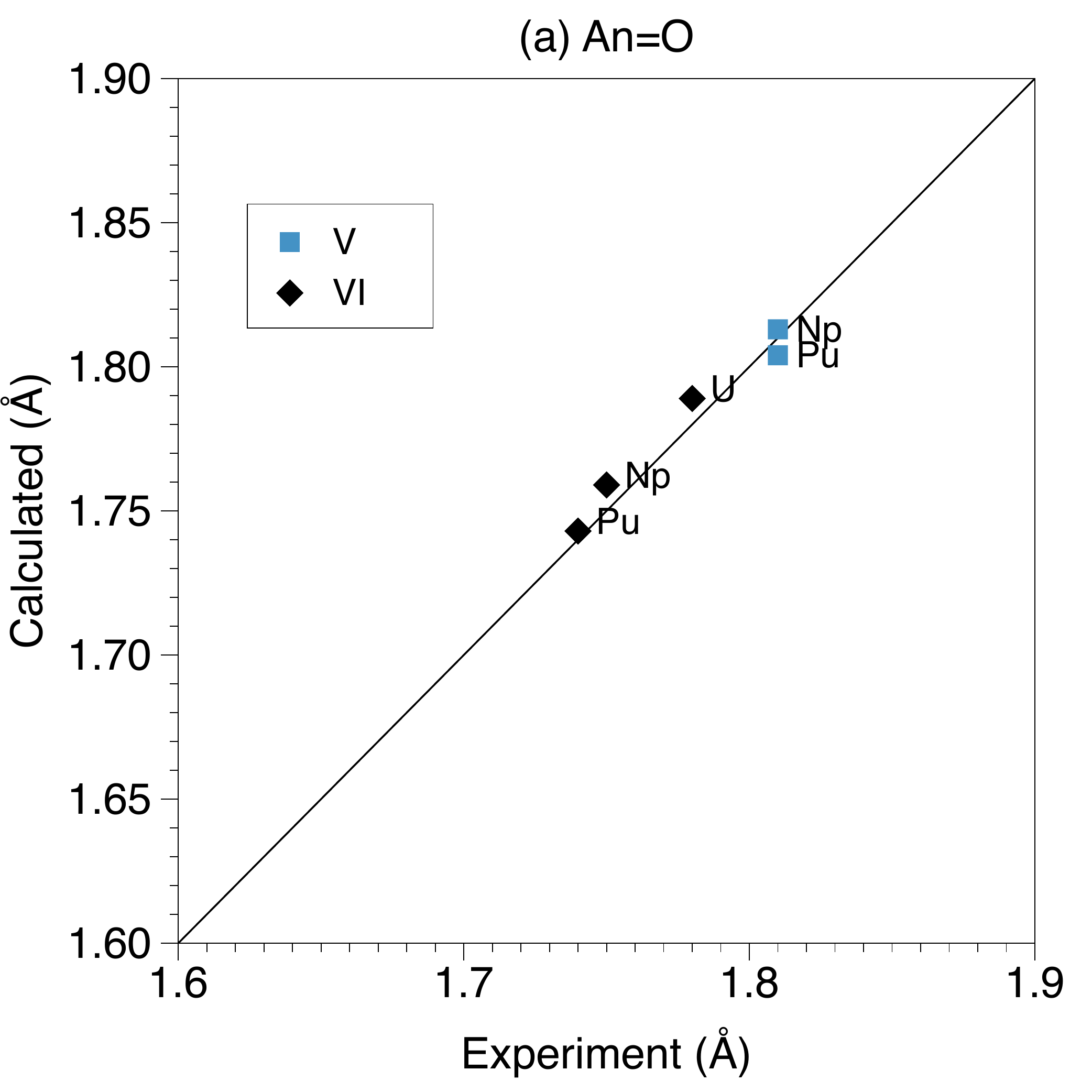} 
\includegraphics[width=80mm]{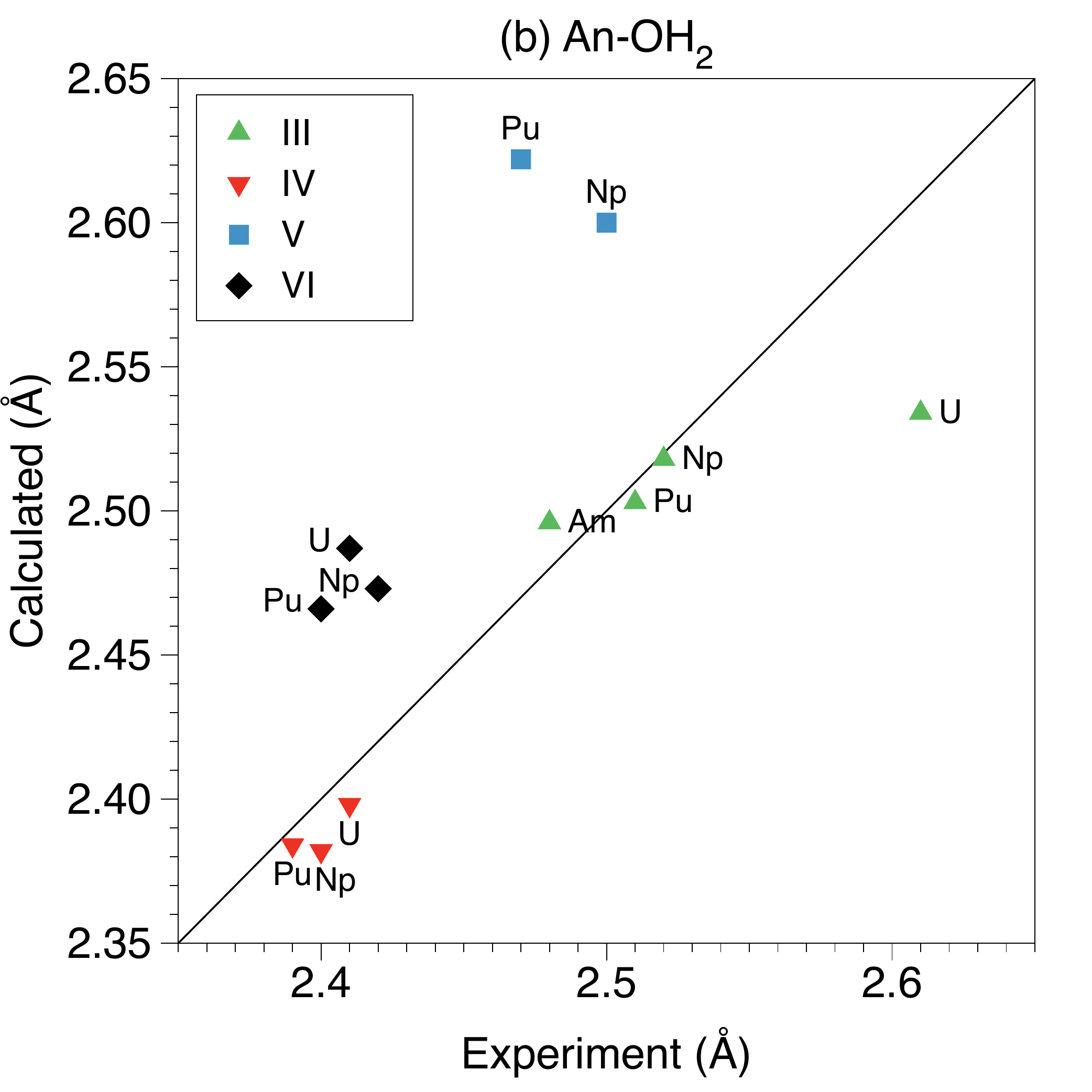} 
\caption{Selected calculated mean internuclear distances compared to
experimental values for  (a) An=O actinyl bonds and (b) An-OH$_2$ bonds between 
the actinide atom and water molecules in the first solvation sphere.}
\label{fig:radii}
\end{figure}

Minimum energy configurations are found to have highly symmetrical
geometries consistent with X-ray absorption fine structure (XAFS) and extended-XAFS (EXAFS) measurements. 
For example, the geometry of
the Pu(III) and Pu(IV) complexes converge to the cubic configuration
shown in Fig. \ref{fig:puGeo} while the actinyls form pentagonal
bipyramids. Interestingly, the converged geometry of the Pu(V) actinyl complex
(Fig. \ref{fig:puGeo} (c)) rotates one of
the water molecules so that both hydrogen atoms are on the same side of the
molecular equatorial plane.  This distortion of the molecular
symmetry is also seen in the converged geometries of U(VI), Np(VI), Pu(V),
Am(V) and Am(VI). Table \ref{tab:bondLengths} shows the calculated
bond lengths of the four oxidation states.   The calculated
geometries compare well with those found in previous DFT
studies\cite{Cao:2005, Hay:2000, Shamov:2005, Sonnenberg:2005,
Tsushima:2001, Tsushima:2005} of actinyls in solution and also, as shown in Fig. \ref{fig:radii},
with available experimental measurements, though DFT overestimates the mean
An-OH$_2$ bond lengths by up to 0.1 \AA~
in the case of the actinyls. 

\begin{table}
\centering
\caption{Calculated and selected experimental internuclear distances (\AA).
The calculated lengths are means over all bonds of the given type.
\label{tab:bondLengths}}
\begin{ruledtabular}
\setlength{\tabcolsep}{1pt}
\begin{tabular*}{0.9 \textwidth}{@{\extracolsep{\fill}}clllllllll}
& & \multicolumn{4}{c}{V} & \multicolumn{4}{c}{VI}\\ \cline{3-6}\cline{7-10}
& & U & Np & Pu & Am    & U & Np & Pu & Am\\
\cline{1-10}
\multirow{2}{*}{An=O$_{\text{eq}}$} & Calc. & 1.84 & 1.81 &
1.80 & 1.80 &  1.79 & 1.76 & 1.74  & 1.74\\
\cline{3-10} & \multirow{3}{*}{Exp.} & & 1.81\footnotemark[1] &1.81\footnotemark[2] & & 1.76\footnotemark[3]& 1.75\footnotemark[4] & 1.74\footnotemark[2] & \\
& & & 1.85\footnotemark[3] & & & 1.78\footnotemark[5] & & \\
& & & & & & 1.702\footnotemark[6] & &  \\
\hline
\multirow{2}{*}{An-OH$_2$} & Calc. & 2.61 & 2.60 & 2.62 & 2.65
& 2.49 & 2.47 & 2.47  & 2.50 \\
\cline{3-10} & \multirow{2}{*}{Exp.} & & 2.47\footnotemark[1] & 2.47\footnotemark[2] &  & 2.41\footnotemark[3]\footnotemark[5] & 2.42\footnotemark[4] & 2.40\footnotemark[2] & \\
& & & 2.50\footnotemark[3] & & & 2.421\footnotemark[6] & & & \\
\cline{1-10} & & \multicolumn{4}{c}{III} & \multicolumn{4}{c}{IV}\\ 
\cline{3-6}\cline{7-10} & & U & Np & Pu & Am  &  U & Np & Pu & \\
\hline
\hline
\multirow{2}{*}{An-OH$_2$} & Calc.  & 2.53 & 2.52 & 2.50 & 2.50 & 2.40 & 2.38 & 2.38 & \\ \cline{3-10}
& \multirow{2}{*}{Exp. }  & 2.61\footnotemark[7] & 2.52\footnotemark[7] & 2.49\footnotemark[2] & 2.48\footnotemark[8]  & 2.41\footnotemark[3] & 2.39\footnotemark[1] & 2.39\footnotemark[2] &\\
& & & 2.51\footnotemark[3]  &    &  & 2.40\footnotemark[3]&  &\\
\end{tabular*}
\end{ruledtabular}
\footnotetext[1]{Reference \onlinecite{Denecke:2005p367}} 
\footnotetext[2]{Reference \onlinecite{Conradson:1998}} 
\footnotetext[3]{Reference \onlinecite{Allen:1997p193}}
\footnotetext[4]{References \onlinecite{Clark:1999} and \onlinecite{Tait:1999} as cited in Reference \onlinecite{Hay:2000}} 
\footnotetext[5]{Reference \onlinecite{Wahlgren:1999}}
\footnotetext[6]{Reference \onlinecite{Aaberg:1983p366}}
\footnotetext[7]{Reference \onlinecite{Cotton:2006}} 
\footnotetext[8]{Reference \onlinecite{Allen:2000p369}} 
\end{table}

\begin{figure}
\begin{center}
\includegraphics[width=120mm]{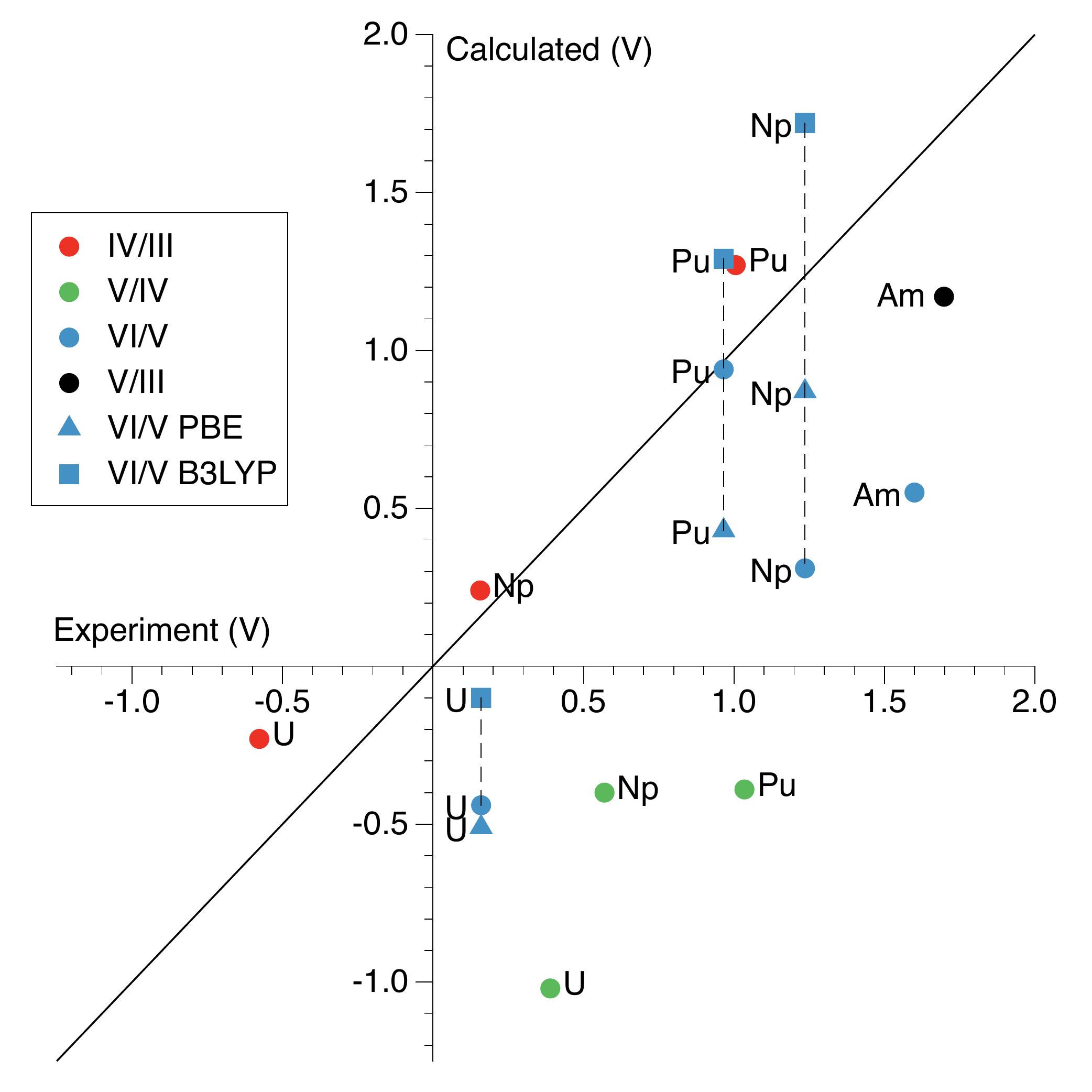}
\end{center}
\caption{Reduction potentials (volts) as calculated by
DFT and corrected for the spin-orbit interaction compared to measured
values as detailed in Ref. \onlinecite{Bratsch:1989}; see however Ref. \onlinecite{Kihara:1999p333} for a different
compilation of measured redox potentials with slightly different values.
PBE and  B3LYP DFT results, corrected for spin-orbit and 
multiplet interactions, are from Table 10 of 
Ref. \onlinecite{Shamov:2005}.}
\label{fig:redoxDFT}
\end{figure}

To calculate the redox potentials we follow the procedure outlined in
Refs. \onlinecite{Hay:2000} and \onlinecite{Shamov:2005}, extending
that work to include the An(V)/An(IV) and An(IV)/An(III) redox couples
as well as reactions involving americium.  The three half-reactions
are as follows: 
\begin{eqnarray}
\text{AnO}_2\text{(H}_2\text{O)}_5^{2+}(\text{VI}) + \text{e}^- &\rightarrow& \text{AnO}_2\text{(H}_2\text{O)}_5^{+}(\text{V})
\nonumber \\
\text{AnO}_2\text{(H}_2\text{O)}_5^{+}(\text{V}) + 4 \text{H}_3\text{O}^+ + \text{e}^- &\rightarrow& 
\text{An}\text{(H}_2\text{O)}_8^{4+}(\text{IV}) + 3 \text{H}_2\text{O}
\nonumber \\
\text{An}\text{(H}_2\text{O)}_8^{4+}(\text{IV}) + \text{e}^- &\rightarrow& \text{An}\text{(H}_2\text{O)}_8^{3+}(\text{III})\ .
\label{eq:halfReactions}
\end{eqnarray}
Potentials are obtained from the free energies of the following three full reactions:
\begin{eqnarray}
\text{AnO}_2\text{(H}_2\text{O)}_5^{2+}(\text{VI}) + \frac{1}{2} \text{H}_2(g) + \text{H}_2\text{O} &\rightarrow&
\text{AnO}_2\text{(H}_2\text{O)}_5^{+}(\text{V}) + \text{H}_3\text{O}^+
\nonumber \\
\text{AnO}_2\text{(H}_2\text{O)}_5^{+} (\text{V})+  3 \text{H}_3\text{O}^+ + \frac{1}{2} \text{H}_2(g) &\rightarrow&
\text{An}\text{(H}_2\text{O)}_8^{4+}(\text{IV}) + 2 \text{H}_2\text{O}
\nonumber \\
\text{An}\text{(H}_2\text{O)}_8^{4+}(\text{IV}) + \frac{1}{2} \text{H}_2(g) + \text{H}_2\text{O} &\rightarrow& 
\text{An}\text{(H}_2\text{O)}_8^{3+}(\text{III})+ \text{H}_3\text{O}^+\ 
\label{eq:reactions}
\end{eqnarray}
using the zero-potential reference half-reaction of the standard hydrogen
electrode (SHE) $\text{H}^+ + \text{e}^- \rightarrow  \frac{1}{2} \text{H}_2(\text{g})$.  
We note that the free energy of the hydronium reaction $\text{H}_3\text{O}^+  \rightarrow \text{H}^+ +
\text{H}_2\text{O}$ as calculated within DFT is $-5.33$ eV. 

Because Am(IV) is unstable, in the case of americium the V/III redox potential is calculated 
in lieu the V/IV and IV/III redox couples. In this case, the reduction half reaction is: 
\begin{eqnarray}
\text{AmO}_2\text{(H}_2\text{O)}_5^{+}(\text{V}) + 4 \text{H}_3\text{O}^+ + 2\text{e}^- &\rightarrow& 
\text{Am}\text{(H}_2\text{O)}_8^{3+}(\text{III}) + 3 \text{H}_2\text{O}
\label{eq:AmhalfReactions}
\end{eqnarray}
corresponding to the full reaction
\begin{eqnarray}
\text{AmO}_2\text{(H}_2\text{O)}_5^{+} (\text{V})+  2 \text{H}_3\text{O}^+ + \text{H}_2(g) &\rightarrow&
\text{Am}\text{(H}_2\text{O)}_8^{3+}(\text{III}) + \text{H}_2\text{O}\ .
\label{eq:Amreactions}
\end{eqnarray}
Fig. \ref{fig:redoxDFT}
compares the calculated potentials to experiment; the calculated potentials include the contribution from
the one-particle spin-orbit interaction as described below in
Sec. \ref{sec:independentParticle}.  The VI/V potentials for U, Np, and Pu are qualitatively
consistent with previous work despite differences in the DFT
packages.  We find redox  potentials, including the
spin-orbit interaction, of respectively $-0.44$, $0.31$ and $0.94$
volts. These compare to values  $2.37$, $4.00$ and $3.28$ volts found by Hay, Martin, and Schreckenbach\cite{Hay:2000} 
who used Gaussian 98, relativistic core potentials, and the hybrid B3LYP
functional, and included multiplet interactions corrections.   However, Shamov and
Schreckenbach\cite{Shamov:2005,Shamov:2006p12072} following a similar procedure but with
a smaller ($n \leq 4$ frozen core obtained  $-0.10$, $1.72$ and $1.29$
volts in better agreement with experiment ($0.160$, $1.236$ and $0.966$
volts)\cite{Bratsch:1989}, highlighting the importance of treating all
orbitals with $n > 4$ dynamically.  Use of the Priroda
PBE functional yielded comparable redox potentials of
$-0.51$, $0.87$ and $0.43$ volts.\cite{Shamov:2005}   When the multiplet interaction correction
is removed these VI/V potentials they show the same trend 
as our calculated DFT potentials , increasing monotonically from U to Np to Pu. 

As mentioned above we only study the case of  
first solvation spheres of An(III) and An(IV) with 8 water molecules.  
Ref. \onlinecite{Clark:2006p427} reports alternative molecular geometries for plutonium ions,
with Pu(III) surrounded instead by 9 coordinating water molecules 
(see also Refs. \onlinecite{Blaudeau:1999p463} and \onlinecite{Tsushima:2005}).  
A DFT calculation of the free energy of Pu(III) with nine water molecules
finds it to be lower by $0.09$ eV, a small change in comparison
to the other corrections that we consider here.   

\section{Independent Particle Model}
\label{sec:independentParticle}

\begin{figure}
\centering
\includegraphics[width=110mm]{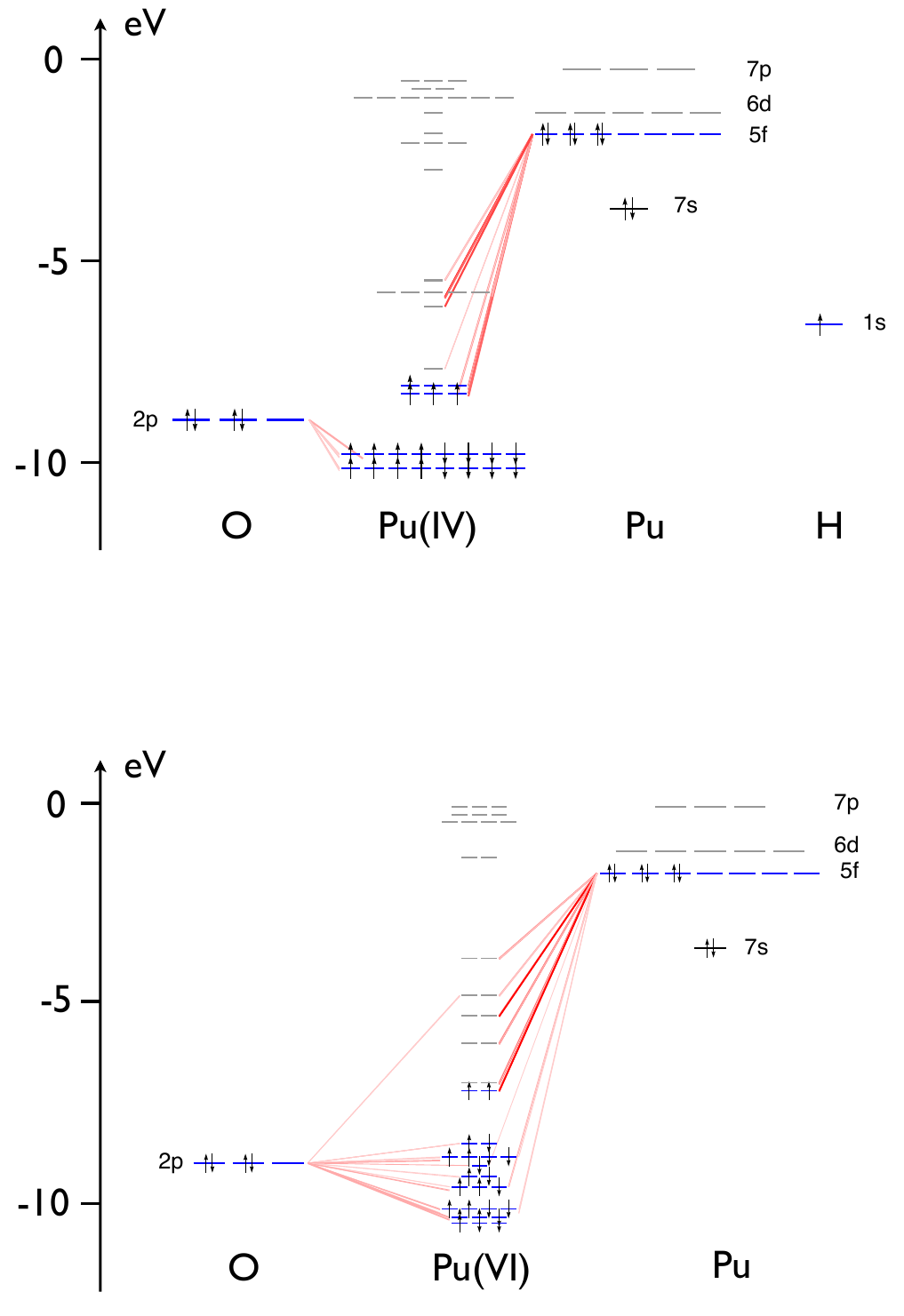} 
\caption{KS molecular orbital energy levels and their
atomic fragment composition for the Pu(IV) (top) and Pu(VI) (bottom) complexes, 
highlighting the importance of the 5f and axial (actinyl) 2p orbitals near the HOMO/LUMO 
boundary.  The relative size of the contributions from the L\"owdin atomic orbitals is indicated by the thickness of the lines.
(Hydrogen orbitals are not shown for Pu(VI) and O includes contributions from both
the actinyl ligands and the water molecules.)
As indicated the atomic fragments have equal numbers of up and down electrons, but the complex has greater 
numbers of up electrons, as the spin-unrestricted DFT calculations satisfy Hund's rule.}
\label{fig:puLevels}
\end{figure}

We turn next to the construction of the independent-particle model of
the frontier orbitals that includes the spin-orbit interaction and
serves as the starting point for the many-body Anderson impurity model: 
\begin{equation}
\hat{H}_{indep} = \hat{H}_{ks} + \hat{H}_{so}\ .
\label{eq:indepH}
\end{equation}
We make the usual assumption that KS fermions may be regarded as physical electrons, 
when in fact the connection between the two is subtle:  The DFT ground state is a Slater determinant of 
KS fermions whereas the electron wavefunction is not generally described by a single determinant.  
Conveniently ADF expresses the KS eigenstates as linear
combinations of localized orbitals orthonormalized by the L\"owdin
procedure\cite{Lowdin:1950}.  These Slater-type orbitals, expressed in terms of Cartesian spherical harmonics, form the basis that we work in.   
In this way the KS orbitals are projected onto the Hilbert subspace that retains only the
actinide 5f L\"owdin orbitals and, in the case of the actinyls, 2p L\"owdin orbitals on the
two oxygen ligands.   As Fig. \ref{fig:puLevels} shows, these
are the atomic orbitals that are the most important constituents of the KS orbitals near the HOMO/LUMO
boundary.    For the actinyls, $14 + 6 + 6 = 26$ L\"owdin orbitals are thus retained; for
oxidation states III and IV only the 14 5f orbitals are required.
Although it would be desirable to also retain the An 6d L\"owdin orbitals
which are close in energy to the An 5f orbitals, and also the L\"owdin orbitals in the first solvation sphere 
of water molecules, the resulting many-body Hilbert space would be too large for exact diagonalizations to be carried out.  

In the case of the actinyls the resulting effective Hamiltonian of the
frontier orbitals may be written: 
\begin{eqnarray}
\hat{H}_{ks} = \sum_{i j \sigma} \epsilon^f_{i j \sigma} f^\dagger_{i \sigma} f_{j \sigma}
+ \sum_{a b \sigma} \epsilon^p_{a b \sigma} p^\dagger_{a \sigma} p_{b \sigma}
+ \sum_{i a \sigma} (t_{i a \sigma} f^\dagger_{i \sigma} p_{a \sigma} + H.c.)\ .
\label{eq:ks}
\end{eqnarray}
Operator $f^\dagger_{i \sigma}$ ($p^\dagger_{a \sigma}$) creates an
electron in the 5f (2p) L\"owdin orbital with spin $\sigma$ and spatial state
$i$ ($a$) in the Cartesian spherical harmonic basis.
In Eq. \ref{eq:ks}  $\epsilon^f_{i j \sigma}$ and $\epsilon^p_{a b \sigma}$ are
matrix elements for, respectively, the An 5f and O 2p L\"owdin orbitals, and $t_{i a
  \sigma}$ are the hopping amplitudes between the 5f and 2p orbitals.
Oxidation states III are IV are modeled by the first term in Eq. \ref{eq:ks} alone.   
Parameters $\epsilon$ and $t$ are obtained by calculating the matrix elements of the  
KS Hamiltonian, which is diagonal in the basis of KS orbitals, 
in the basis of the 5f (and in the case of actinyls, 2p) L\"owdin orbitals.  The matrix elements
may then be grouped into the amplitudes that appear in  Eq. \ref{eq:ks}.

\begin{figure}
\begin{center}
\includegraphics[width=160mm]{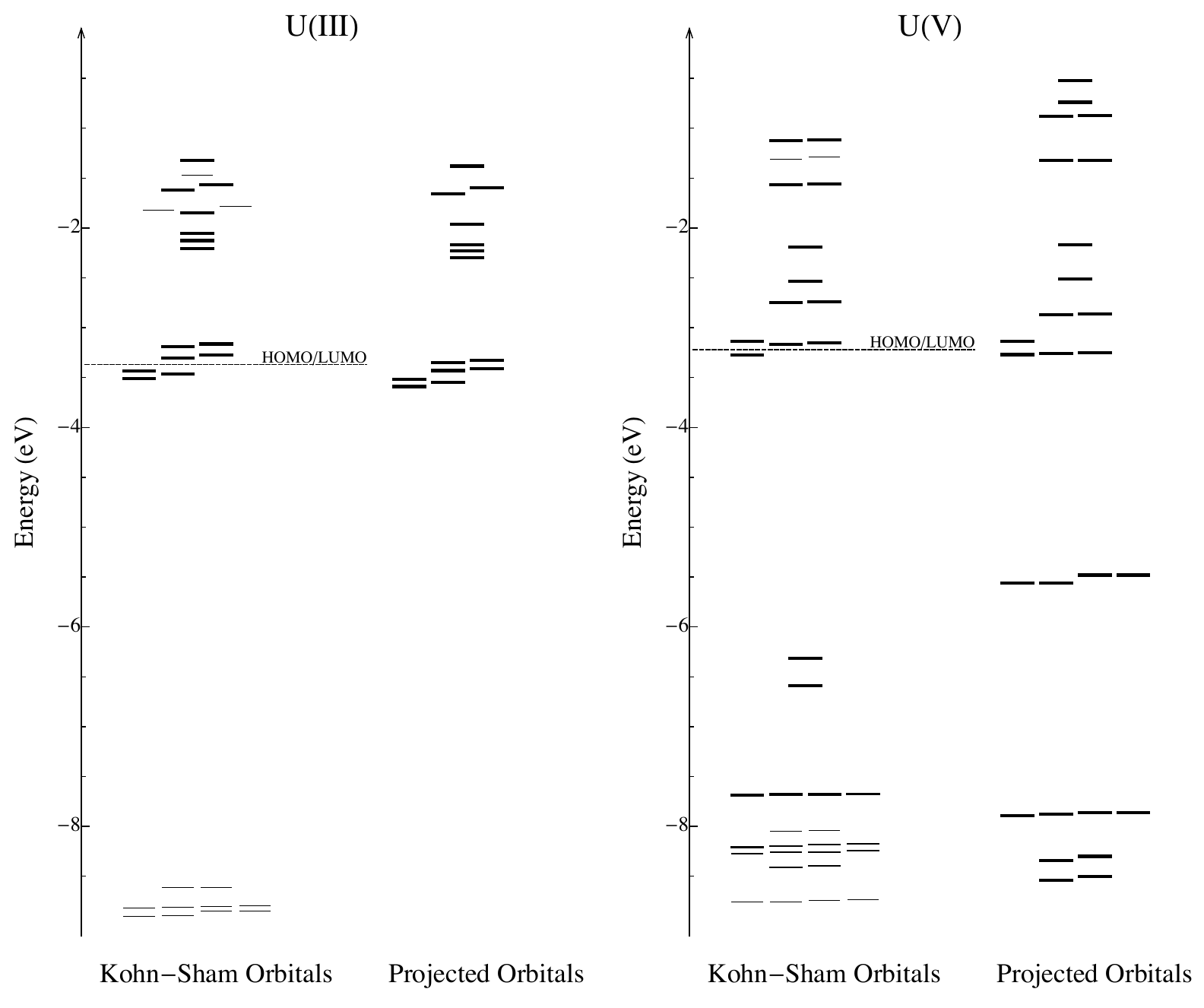}
\end{center}
\caption{Comparison of the KS molecular orbital energies against those of the independent-particle model
that describes the projected 5f and 5f-2p subspaces, for the cases of U(III) and U(V).  Shading of the KS orbitals 
indicate the relative size of their contribution to the 5f or 5f-2p L\"owdin subspace, with the lightest lines representing orbitals that make little
or no contribution.}
\label{fig:match}
\end{figure}

\begin{figure}
\begin{center}
\includegraphics[width=120mm]{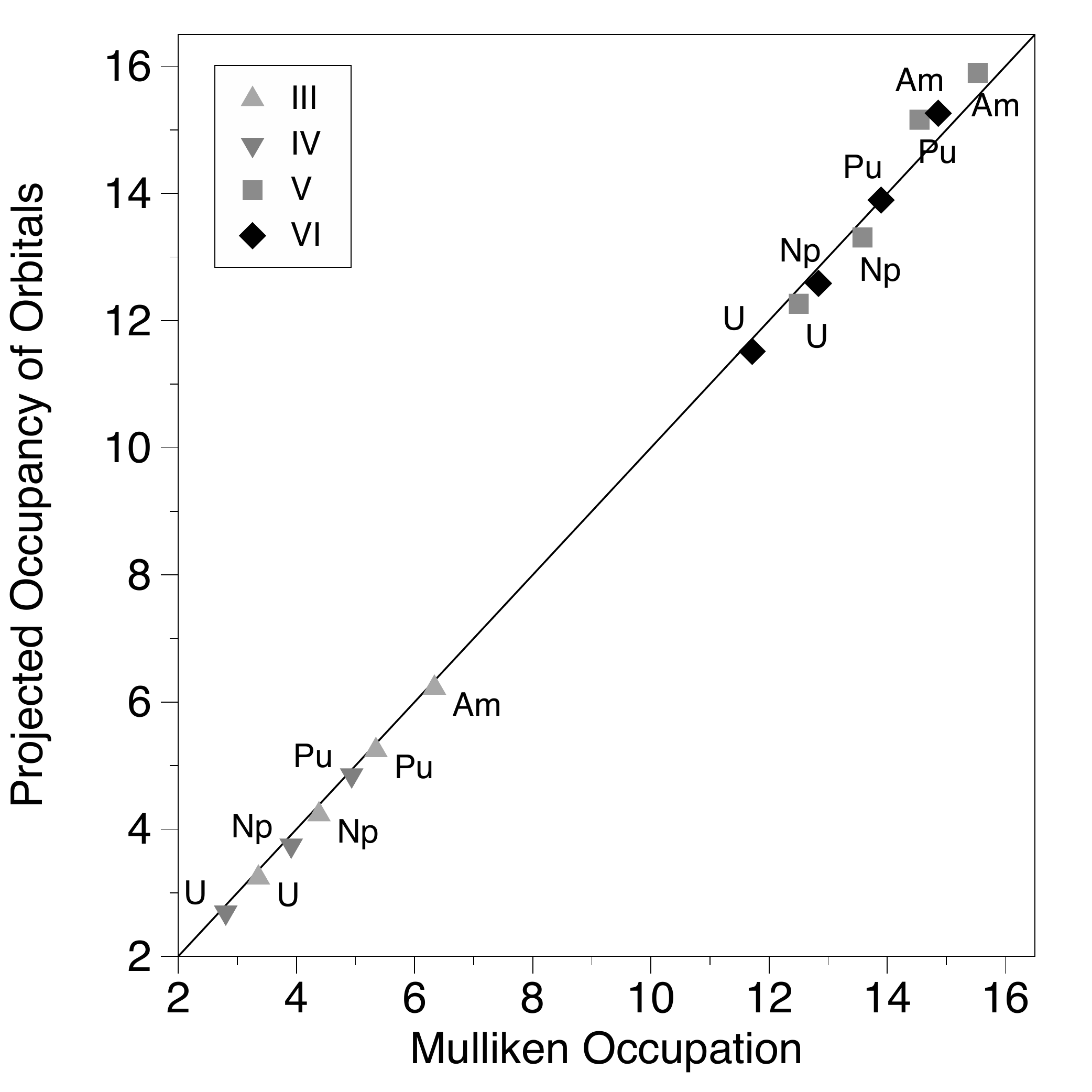} 
\end{center}
\caption{Total electronic occupation of the An 5f and (in the case of actinyls) O 2p L\"owdin orbitals 
as calculated by projection of the occupied KS molecular orbitals onto 
the L\"owdin atomic orbitals, compared to that found by Mulliken analysis.} 
\label{fig:mull}
\end{figure}

The truncation of the full Hilbert space to the subspace consisting only of An 5f and (actinyl)
O 2p L\"owdin orbitals introduces error into the calculations.  Fig. \ref{fig:match} shows how the single-particle 
energies of the projected independent-particle model match those of the full unprojected KS
levels for the cases of U(III) and U(V).  
In Fig. \ref{fig:mull} the total electronic occupation of the An 5f and (actinyl) O 2p L\"owdin orbitals 
as calculated by projection of the occupied KS molecular orbitals onto 
the L\"owdin atomic orbitals is compared to the corresponding occupancies found by Mulliken analysis.
The occupancy of oxidation states III and IV agree to within 2\% of 
that found by Mulliken analysis; and to within 4\% for oxidation states V and VI.  

As it stands Eq. \ref{eq:indepH} has no spin-flip processes, reflecting the limitation of DFT 
which is formulated in terms of separate densities of spin-up and down electrons.   
The one-electron spin-orbit interaction for the 5f L\"owdin orbitals,  
$\hat{H}_{so} = \zeta \vec{\ell} \cdot \vec{s} + \zeta_0$, is therefore added to the independent-particle
model.  Energies $\zeta$ and $\zeta_0$ are
obtained from ADF calculations on isolated (gas-phase) actinide ions.
The spin-orbit interaction splits the energies of the $j = 5/2$ and
$7/2$ states by  $\Delta \epsilon = \epsilon_{7/2} - \epsilon_{5/2} = \frac{7}{2} \zeta
= 0.67$ eV (U), $0.78$ eV (Np), $0.87$ eV (Pu) and $1.06$ eV (Am).  The
interaction also shifts the energy of all 5f L\"owdin orbitals upwards in energy by $\zeta_0 = 0.18$ eV
(U), $0.23$ eV (Np), $0.26$ eV (Pu) and $0.29$ eV (Am).   

\section{Many-Body Model}
\label{sec:manyBody}
The Coulomb repulsion between 5f electrons is taken into account by adding the corresponding two-body interaction term to the Hamiltonian.  The two-body interaction
\begin{equation}
\hat{H}_{ee} = \frac{1}{2} \sum_{m_1 m_2 m_3 m_4 \atop{m_1+m_2 = m_3+m_4}}
I_{m_1, m_2, m_3, m_4}~ f^\dagger _{m_1 \sigma_1}
f^\dagger_{m_2 \sigma_2} f_{m_3 \sigma_2} f_{m_4 \sigma_1}  
\label{eq:Coulomb}
\end{equation}
is conveniently represented in terms of the Coulomb matrix elements
\begin{equation}
I_{m_1, m_2, m_3,m_4} =  \sum_{L=0, 2, 4, 6}
F^L c^{(L)}(m_1,m_4)c^{(L)}(m_3,m_2)  
\end{equation}
where $L$ is the total orbital angular momentum of two 5f electrons, $c^{(L)}$ are the Gaunt coefficients\cite{Gaunt:1929,RacahII:1942} and $F^L$ are the Slater integrals.\cite{Slater:1929,Condon:1931}   As discussed below in Sec. \ref{sec:results} the $F^2$, $F^4$, and $F^6$ integrals parameterize the energetics of rearrangements of the electrons in the 5f shell and hence may be 
determined from spectroscopic data.  $F^0$, however, is sensitive only to the total number of electrons in the 5f shell and is expected to be highly screened.  We treat it is the one adjustable parameter in the hybrid calculation. 
Like $\hat{H}_{indep}$, the electrostatic repulsion terms are transformed into the Cartesian spherical harmonic basis for the purpose of numerical calculations.  We performed numerical tests to verify that the spectrum of Eq. \ref{eq:Coulomb} reproduces published results for isolated actinide atoms\cite{Norman:1995}. 

As the interaction is already partially included at the DFT level, care must be taken to avoid double counting it.\cite{Albers:2009p110}  
This we do by subtracting its contribution at the HF level of approximation, making an assumption, however, that DFT with the PBE
functional is close to HF in its treatment of the interaction.  Thus 
\begin{eqnarray}
\hat{H} = \hat{H}_{indep} + \hat{H}_{ee} - \overline{H}_{ee}
\label{eq:fullH}
\end{eqnarray}
where the overline denotes the HF factorization of $\hat{H}_{ee}$.  
The one-body HF subtraction, $\overline{H}_{ee}$, that models the part of the Coulomb interaction already included in the DFT calculation is given by
\begin{equation}
\label{eq:HF}
\overline{H}_{ee} = \frac{1}{2}\sum_{m, m^\prime, \sigma} \left(J_{m m^\prime} - K_{m m^\prime \sigma} \right) f^\dagger_{m \sigma} f_{m^\prime \sigma}
\end{equation}
where the direct or Hartree interaction $J$ is given by
\begin{equation}
\label{eq:hfJ}
J_{m m^\prime} = \sum_{n, n^\prime, \sigma} I_{m,n, n^\prime, m^\prime}~ \langle f^\dagger_{n \sigma} f_{n^\prime \sigma} \rangle
\label{eq:J}
\end{equation}
and the exchange or Fock interaction K is 
\begin{equation}
\label{eq:hfK}
K_{m m^\prime \sigma} = \sum_{n, n^\prime} I_{m,n, m^\prime, n^\prime}~ \langle f^\dagger_{n \sigma} f_{n^\prime \sigma}\rangle.
\label{eq:K}
\end{equation}
The expectation values appearing in Eqs. \ref{eq:J} and \ref{eq:K} are calculated from DFT.   
By construction, then,  $\langle \hat{H}_{ee} - \overline{H}_{ee}
\rangle = 0$ providing a valuable check on the numerical calculations. 

\section{Fractional Occupancy}
\label{sec:fractionalOccupancy}

The combined occupancy of the 5f and 2p L\"owdin orbitals is not  an integer (see Fig. \ref{fig:mull}).  
To handle this fractional occupancy within the many-body model of the reduced 5f-2p subspace, we calculate the ground state
energies of the many-body model at the two integer occupancies that bracket the fractional value, and
then compute a weighted average of the energies.  
The effectiveness of this algorithm may be illustrated with a simple model consisting of a single f-orbital and a single c-orbital, where the c-orbital is a model for orbitals not included in the restricted 5f-2p subspace.  The model is parameterized by on-site f-orbital energy $\epsilon$, a hopping amplitude between the f- and c-orbitals $t$, and the Coulomb repulsion $U$ between two electrons in the f-orbital:
\begin{eqnarray}
\hat{H} = \epsilon~ f^\dagger_\sigma~ f_\sigma - t~ (f^\dagger_\sigma~ c_\sigma + H.c.) 
+ U~ f^\dagger_\uparrow f^\dagger_\downarrow f_\downarrow f_\uparrow
\end{eqnarray}
where the sum over the repeated $\sigma$ spin index is implied. 
The model is easily diagonalized.  For instance in the 2-particle, spin-singlet, subspace spanned by the 3 basis vectors:
\begin{eqnarray}
| 1 \rangle &\equiv& c^\dagger_\uparrow~ c^\dagger_\downarrow~ | 0 \rangle
\nonumber \\
| 2 \rangle &\equiv& \frac{1}{\sqrt{2}} (c^\dagger_\uparrow~ f^\dagger_\downarrow - c^\dagger_\downarrow~ f^\dagger_\uparrow) | 0 \rangle
\nonumber \\
| 3 \rangle &\equiv&  f^\dagger_\uparrow~ f^\dagger_\downarrow | 0 \rangle~ 
\end{eqnarray}
$\hat{H}$ takes the form of a $3 \times 3$ matrix and, when diagonalized, the resulting exact ground state energy $E_0$ may be compared against approximations.

The HF approximation to the Hubbard interaction, $\hat{H}_{ee} = U f^\dagger_\uparrow  f^\dagger_\downarrow f_\downarrow f_\uparrow = U n_\uparrow n_\downarrow$, is given by setting $I = U$ in Eq. \ref{eq:HF} and 
using the fact that $\langle n_\uparrow \rangle = \langle n_\downarrow \rangle = \frac{1}{2} \langle n \rangle$ by spin-rotational invariance of the spin-singlet ground state where $n \equiv n_\uparrow + n_\downarrow = f^\dagger_\uparrow f_\uparrow + f^\dagger_\downarrow f_\downarrow$.  The result is:
\begin{eqnarray}
\overline{H}_{ee} = \frac{U}{4} \langle n \rangle~ n 
\end{eqnarray}
and it replaces the two-body interaction with one-body term that renormalizes the one-body f-orbital energy:
\begin{eqnarray}
\epsilon \rightarrow \epsilon_r(\langle n \rangle) = \epsilon + \frac{U}{4}  \langle n \rangle\ .
\end{eqnarray}
Self-consistency is then attained by adjusting $\langle n \rangle$ so that the f-orbital occupancy as calculated in the ground state of the independent-particle Hamiltonian $\hat{h}$:
\begin{eqnarray}
\hat{h} = \epsilon_r(\langle n \rangle)~ f^\dagger_\sigma~ f_\sigma - t~ (f^\dagger_\sigma~ c_\sigma + H.c.) 
\end{eqnarray}
equals $\langle n \rangle$.  The resulting HF equation is
\begin{eqnarray}
\langle n \rangle = \frac{2}{1 + \left(\frac{\epsilon_r(\langle n \rangle) + \sqrt{\epsilon_r^2(\langle n \rangle) + 4 t^2}}{2 t}\right)^2}
\end{eqnarray}
and it can be solved by iteration.  The HF approximation to the ground-state energy of the two-electron system is then given by:
\begin{eqnarray}
\overline{E}_0 = \epsilon_r(\langle n \rangle) - \sqrt{\epsilon_r^2(\langle n \rangle) + 4 t^2} 
\end{eqnarray}
which is simply the energy from filling the lowest eigenstate of the renormalized independent-particle model with both a spin-up and a spin-down electron.  

An improved approximation of the ground state energy can be obtained by carrying out an exact diagonalization in the reduced subspace consisting of only the f-orbital.   Double-counting the interaction is avoided by subtracting the HF contribution to the Coulomb energy, $\overline{H}_{ee}$ from the exact two-body Hubbard term $U n_\uparrow n_\downarrow = (U/2)(n^2 - n)$.  The improved estimate of the ground state energy is given, for fixed integer occupancy $n = 0, 1, 2$ of the f-orbital, by Eq. \ref{eq:fullH} which in this simplified context reads:
\begin{eqnarray}
\tilde{E}_0(n) = \overline{E}_0 + \frac{U}{2} (n^2 - n) - \frac{U}{4} \langle n \rangle~ n .
\end{eqnarray}
Finally a weighted average of  $\tilde{E}_0(n)$ based on the HF occupancy $\langle n \rangle = 1 + x$ yields, for $x > 0$:
\begin{eqnarray}
\tilde{E}_0 = (1 - x)~ \tilde{E}_0(1) + x~ \tilde{E}_0(2)
\end{eqnarray}
and as Table \ref{tab:fractionalOccupancyComparison} shows, for the case of $t = 1$ and $\epsilon = -3$ there is a substantial improvement over the HF approximation.  The ground state energy decreases because the two electrons are now correlated and able to avoid each other.

\begin{table}
\caption{Comparison of ground state energies for the two-orbital model with $t = 1$ and $\epsilon = -3$. Here  $\langle n \rangle$ is the occupancy of the f-orbital in the HF approximation; 
$\overline{E}_0$ is the HF energy; $\tilde{E}_0$ is the improved estimate of the ground state energy; and $E_0$ is the exact ground state energy.}
\begin{ruledtabular}
\begin{tabular}{l c l c l c l c | c |}
$U$ & $\langle n \rangle$  & $\overline{E}_0$ & $\tilde{E}_0$ & $E_0$\\
 \hline
4 & 1.579 & -3.874 & -4.051 & -4.323\\
6 & 1.407 & -3.079 & -3.607 & -4\\
8 & 1.246 & -2.571 & -3.708 & -3.860\\
\end{tabular}
\end{ruledtabular}
\label{tab:fractionalOccupancyComparison}
\end{table}

\section{Results}
\label{sec:results}

The Slater integrals $F^2$, $F^4$, and $F^6$ parameterize changes in electrostatic
energy due to rearrangements of the electrons in the 5f shell, and as a consequence are insensitive 
to the chemical environment surrounding the actinide. We use the values displayed in
Table \ref{tab:ActSpect}.  These are based upon spectroscopic data\cite{Veal:1977p7} that is 
expressed in terms of Racah parameters $E^1$, $E^2$, and $E^3$.  The Racah parameters 
are linearly related to the Slater integrals by the following equations:
\begin{eqnarray}
 F^2 &=& \frac{43}{3} E^1 + \frac{6895}{9} E^2 + \frac{530}{9} E^3
 \nonumber \\
 F^4 &=& \frac{99}{7} E^1 - \frac{12870}{7} E^2 + \frac{396}{7} E^3
 \nonumber \\
F^6 &=& \frac{143}{9} E^1 + \frac{3904}{7} E^2 - \frac{1004}{9} E^3\ .
\end{eqnarray}
The values for $F^4$ and $F^6$ listed in Table  \ref{tab:ActSpect} are somewhat lower than those obtained from Ref. \onlinecite{Veal:1977p7}
(a range of values may be found in the literature, see for instance Ref. \onlinecite{Norman:1995}) but 
we have checked that the corrections to the redox potentials are insensitive to these differences.  
By contrast $F^0$ parameterizes the part of the Coulomb energy that is sensitive to the total number of electrons occupying the 5f shell and is  thus important in charge-transfer reactions.  As expected the corrections to the redox potentials vary with $F^0$ (see below).  As discussed above in 
Sec. \ref{sec:manyBody} $F^0$ is 
expected to be highly screened, but the degree of screening is difficult to predict reliably from
first-principles.   We therefore treat it as the one adjustable parameter in our calculations.  

\begin{table}
\caption{Slater integrals (eV) used in the Anderson model. 
\label{tab:ActSpect}}
\centering
\begin{ruledtabular}
\begin{tabular}{ccccc}
Ion & Configuration & $F^2$ & $F^4$ & $F^6$ \\
\hline
U$^{4+}$ & 5f$^2$ & 5.746&3.693 &2.201 \\
Np$^{4+}$ & 5f$^3$ & 6.249&4.016 &2.394 \\
Pu$^{4+}$ & 5f$^4$ &6.778 &4.356 &2.597 \\
Am$^{4+}$ & 5f$^5$ &7.867 &5.056 &3.014 \\
\end{tabular}
\end{ruledtabular}
\end{table}

\begin{figure}
\begin{center}
\includegraphics[width=120mm]{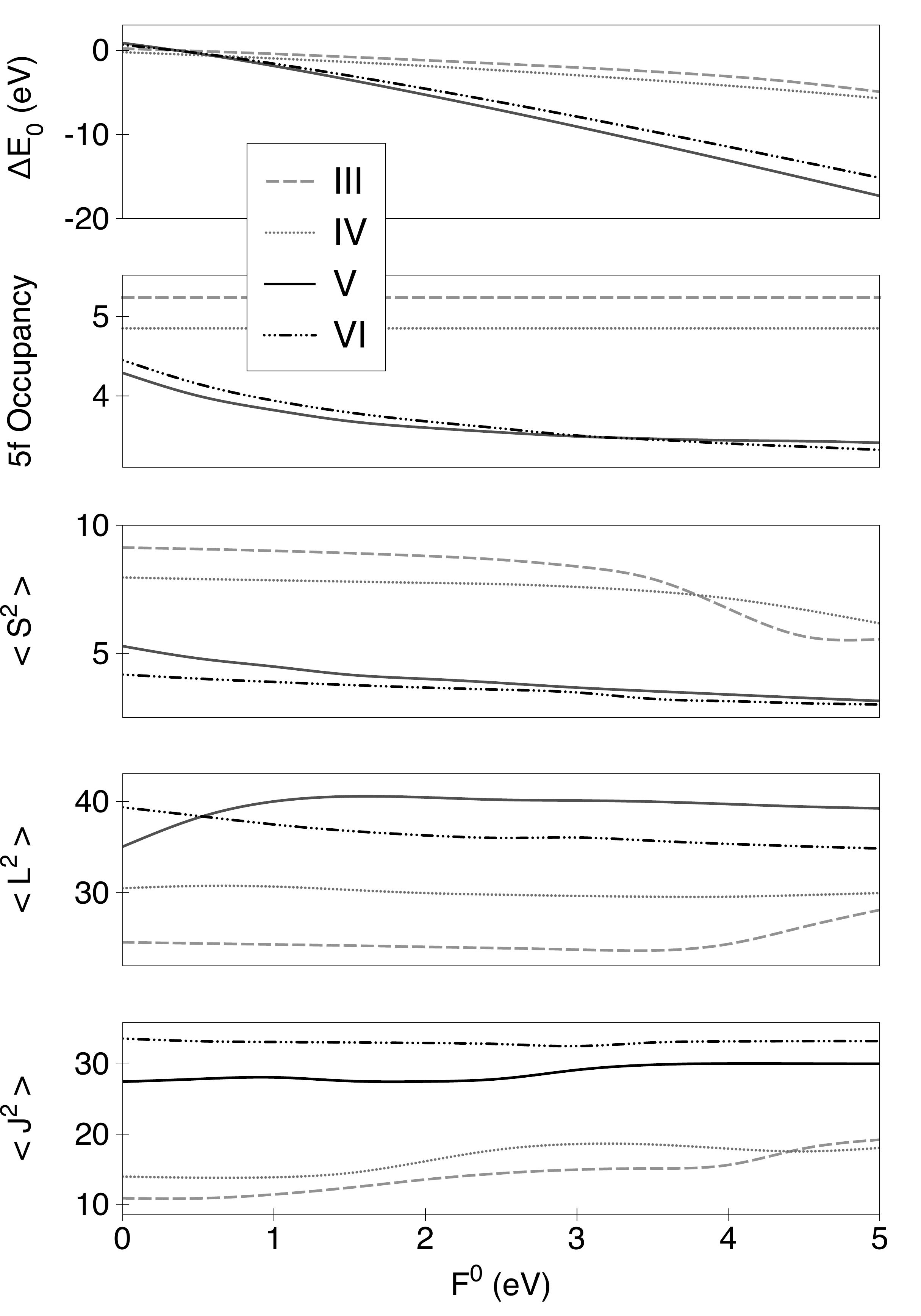} 
\end{center}
\caption{The correction to the DFT ground state energy, the occupancy of the 5f L\"owdin orbitals, the total spin $\langle \vec{S}^2 \rangle$ in the 5f orbitals, the total orbital angular momentum $\langle \vec{L}^2 \rangle$ in the 5f orbitals,
and the total angular momentum $\langle \vec{J}^2 \rangle$ in the 5f orbitals as F$^0$ is varied, holding $F^2$, $F^4$, and $F^6$ and the spin-orbit interaction fixed.  The four oxidation states of plutonium are shown.}
\label{fig:PuAnalysis}
\end{figure}

The many-body Hilbert space has a maximum
dimension $26! / 13!^2 = 10,400,600$ for the case of 13 electrons populating
the 5f and 2p L\"owdin orbitals of an actinyl.  Exact diagonalization of the
many-body Hamiltonian is accomplished with the use of the
sparse-matrix Davidson algorithm.  As the HF approximation may
be formulated as a variational problem over the subset of wavefunctions 
that are single Slater determinants, in the absence of the spin-orbit 
interaction diagonalization of $\hat{H}$, Eq. \ref{eq:fullH}, yields
ground state energies that are less than the ground state energy of
$\hat{H}_{indep}$, Eq. \ref{eq:indepH}.  The reduction in the energy
is a consequence of the fact that correlations between the 5f
electrons permit the electrons to avoid each other more effectively
than when the interaction is described only at the mean-field level.   The effect, 
which holds even in the presence of the spin-orbit interaction, 
is evident in Fig. \ref{fig:PuAnalysis} where it can be seen that 
the electronic ground state energy of the plutonium complexes decreases
with increasing $F^0$.   Also as $F^0$ increases, 
electrons in the actinyls move from the 5f L\"owdin orbitals to the 2p L\"owdin orbitals of the oxygen ligands (the occupancy
remains fixed for Pu(III) and Pu(IV) because only the 5f  orbitals are included in the many-body
model).  This charge transfer is also evident in the total spin in the 5f orbitals which decreases in the
actinyls with increasing $F^0$.  These observables, like the ground state energy,
are calculated from a weighted average of the two many-body ground states of integer occupancies that bracket the occupancy obtained from DFT. 

\begin{figure}
\begin{center}
\includegraphics[width=150mm]{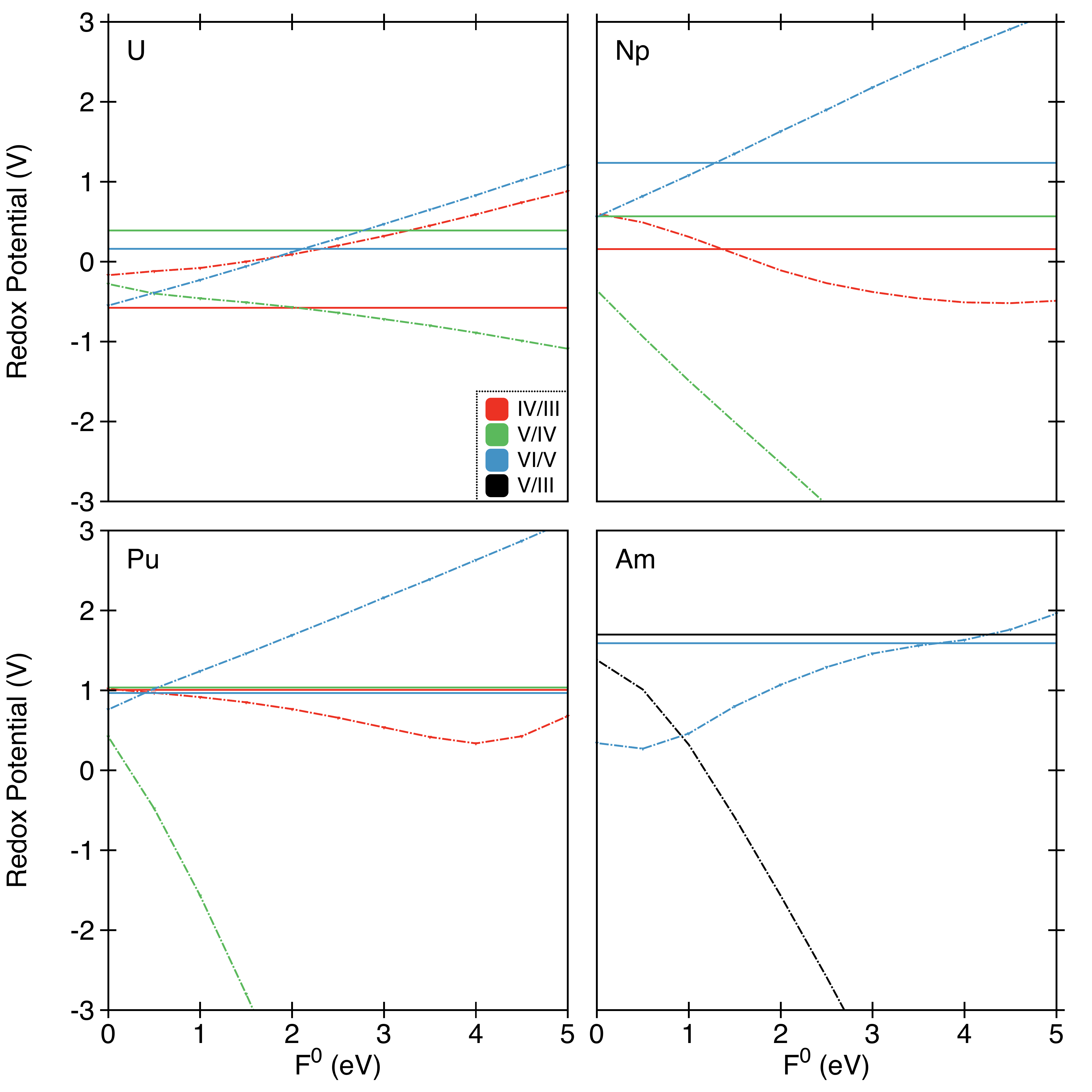}
\end{center}
\caption{Calculated redox potentials of the actinide redox couples  An(IV)/An(III) (red), An(V)/An(IV) (green), An(VI)/An(V) (blue) and Am(V)/Am(III) (black) as a function of  $F^0$. Experimentally determined potentials are indicated by the solid horizontal lines.} 
\label{fig:redoxVeal}
\end{figure}

The many-body correction to the ground
state energy as computed within DFT changes the free energies of the 
reactions Eqs. \ref{eq:reactions} and \ref{eq:Amreactions} and hence the
redox potentials.   Fig. \ref{fig:redoxVeal} shows the
change in the redox potentials as a function of $F^0$. 
Potentials for $F^0 = 0$ are the DFT values corrected by the spin-orbit interaction and the
Slater integrals $F^2$, $F^4$, and $F^6$.  For the VI/V couples (shown
in blue), the closest match with experiment is for $F^0 = 2.1$ eV (U), 
$1.3$ eV (Np), $0.4$ eV (Pu) and $3.7$ eV (Am).  For the
IV/III redox couples (red),  the best match occurs for  $F^0 = 1.4$ eV (Np) and
$0.1$ eV (Pu) eV, but the correction to the electronic energies worsens 
the match to experiment in the case of U.  However, in the case of the V/IV redox
potentials (green), as well as the Am V/III potential (black), the pure DFT values
are all low in comparison to experiment 
(see Fig. \ref{fig:redoxDFT}), and the corrections only lower the potentials further.

\begin{figure}
\begin{center}
\includegraphics[width=120mm]{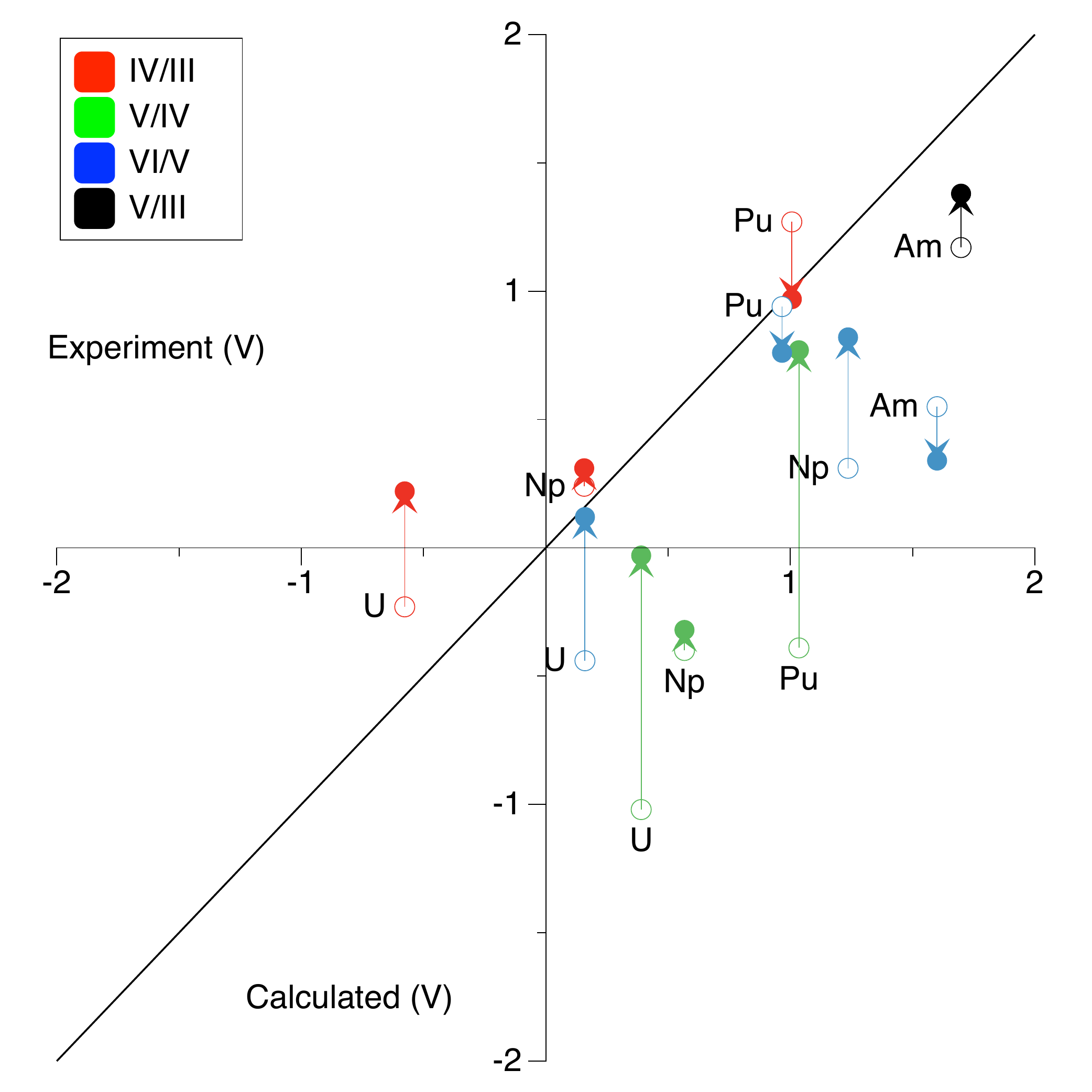}
\end{center}
\caption{Many-body corrected redox potentials (filled symbols) compared
to measured values.  Open symbols  are the potentials obtained from DFT including only the spin-orbit correction.  
$F^0 = 2.5$ eV for U(III) and U(IV); $1.0$ eV for Np(III) and (IV); $0.5$ eV for Pu(III) and (IV);
and $0.5$ eV for Am(III).  In each case $F^0$ is reduced by $0.5$ eV for oxidation states V and VI
to account for increased screening of the Coulomb interaction in the actinyls.} 
\label{fig:redoxVealCorrShift}
\end{figure}

As it stands the calculation does not account for changes in the 
screening of the Coulomb interaction as the oxidation state changes.
A reduction in the size of $F^0$ for the actinyls
yields in most cases a better match with experiment, 
particularly in the case of the V/IV redox couple.  In Fig. \ref{fig:redoxVealCorrShift}
$F^0 = 2.5$ eV for U(III) and U(IV); $1.0$ eV for Np(III) and (IV); $0.5$ eV for Pu(III) and (IV);
and $0.5$ eV for Am(III).  In each case $F^0$ is reduced by $0.5$ eV for oxidation states V and VI.
There is a striking improvement in the match with experiment with two exceptions:
The calculated U IV/III and Am VI/V potentials move further away from the measured values.
The downward shift in $F^0$ may reflect the role that the oxygen ligands 
play in screening repulsion between 5f electrons.   A similar trend has been found in 
studies of isolated molecules and ions.  For example in Ref. \onlinecite{Norman:1995} a range of
values for  $F^0 = 2.3$ to $3.3$ eV are reported for an isolated U$^{4+}$ ion, decreasing to
$F^0 = 1.6$ eV in the case of the UPt$_3$ molecule.

\begin{figure}
\centering
\includegraphics[width=165mm]{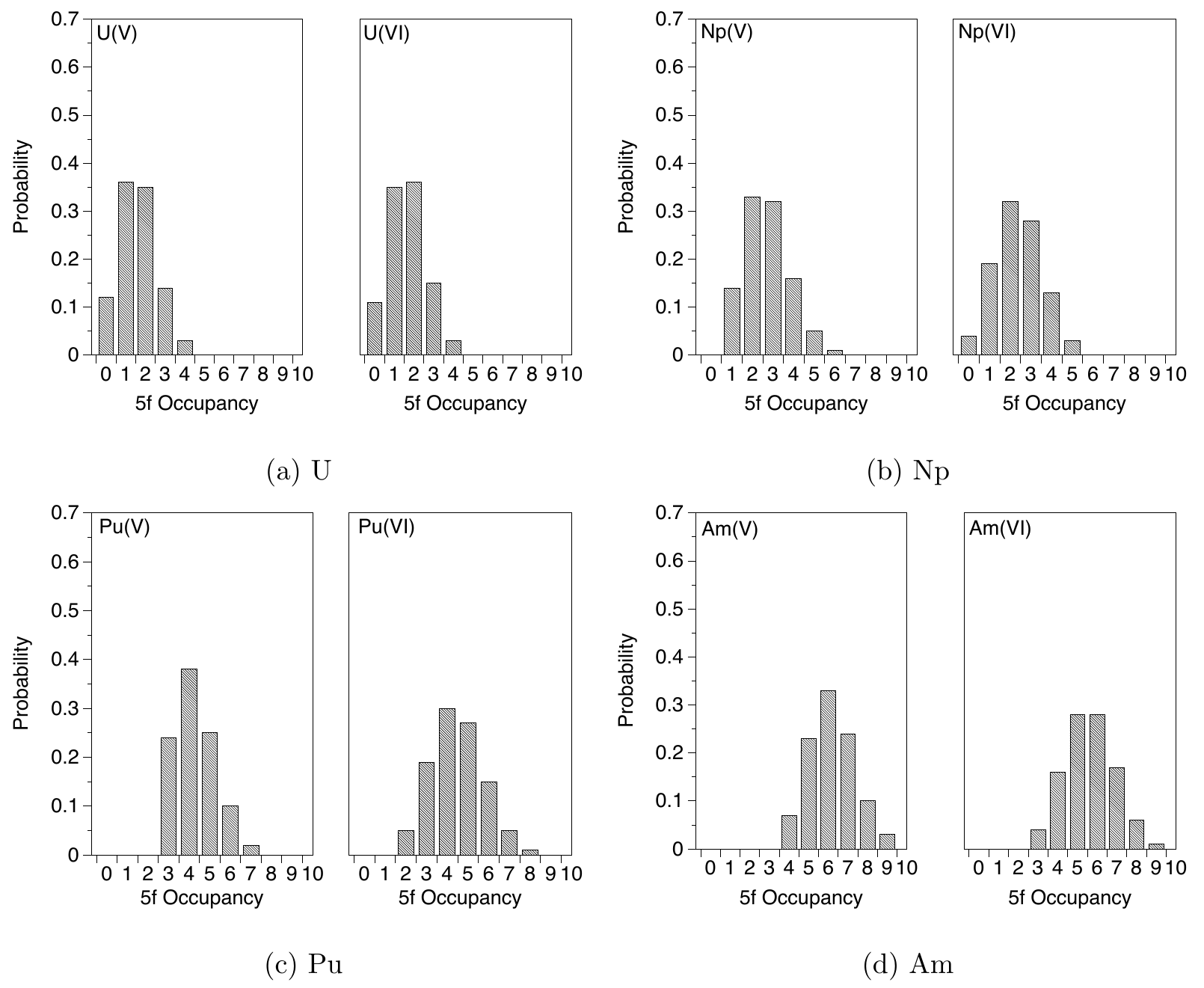} 
\caption{Histogram of the probability for different 5f electronic occupancies for oxidation 
states V and VI.  $F^0$ has the same values as in Fig. \ref{fig:redoxVealCorrShift}.}
\label{fig:Hist}
\end{figure}
Fig. \ref{fig:Hist} presents probability distributions of different electron occupancies\cite{Shim:2007p314,Yee:2010p452}
in the actinyl 5f L\"owdin orbitals, using the same values of $F^0$ adopted in Fig. \ref{fig:redoxVealCorrShift}.
The distributions are particularly broad for neptunium and plutonium reflecting the increasing number of 5f electrons 
as one moves along the row of early actinides competing against an increasing tendency to localize.  

\section{Conclusion}
\label{sec:conclusion}

The hybrid DFT / many-body approach taken in this paper shows some promise for the theoretical
modeling of the difficult but important problem of redox chemistry involving the early actinide elements.  
By incorporating the physics of strong correlations between electrons in the frontier orbitals we are able to
correct the electronic contribution to the free energy as computed in DFT, and thereby 
bring the calculated redox potentials into closer agreement with measured values.  The calculations 
require one adjustable parameter, $F^0$, for each actinide species, in addition to a $0.5$ eV downshift in $F^0$
for the actinyls, yet has predictive power as it yields potentials for 3 redox couples 
An(VI)/An(V), An(V)/An(IV), and An(IV)/An(III).  

The match with experiment is improved in 6 of the 11
redox reactions that we study; agreement remains good in the case of 2 reactions (Pu VI/V and Np IV/III), little changed but poor for
Np V/IV, and 
worsens for 2 others (U IV/III and Am VI/V.  The calculated potentials are certainly not of chemical accuracy, but they do
evidence significant trends.  In the case of plutonium, for instance, the calculated potentials approach the  
measured near-degeneracy of the 3 redox potentials, and hence go some distance towards 
explaining the propensity of plutonium species in solution to easily disproportionate and co-exist in several different oxidation states.\cite{Clark:2000}  In the language of Hubbard models, disproportionation may be viewed
as a consequence of an effective negative-U interaction for the overall complex; see for instance
 Refs. \onlinecite{Watkins:1984,vanderMarel:1988p330,Harrison:2006p316}. In our calculations
its origin may be traced in part to the strong correlations between the 5f electrons that permit the electrons to avoid each other
more efficiently than they can at the level of LDA/GGA or HF, lowering the electronic energy.  When combined with 
all the other contributions to the 
free energy (solvation, vibrations, and translations) the redox potentials become degenerate and
an overall effective negative-U interaction emerges.  

The calculations can be improved or extended in several ways.  At the DFT level, different geometries with varying numbers of 
water molecules in the solvation spheres can be investigated.  It may be desirable to treat a second sphere of 
solvation quantum mechanically rather than with continuum dielectric models.  Of geochemical interest is 
actinide complexation with carbonate and silicate substrates, and with colloids,\cite{Clark:2000,Kubicki:2009p125} and these could be 
investigated by the hybrid approach.    
The projection onto L\"owdin orbitals could be replaced with projective
orthogonalization\cite{Toropova:2007} to better minimize the admixture of 
neglected orbitals.  It may be possible to include some 
additional orbitals in the many-body model, if not by exact diagonalization then possibly by methods such as the 
density-matrix renormalization-group (DMRG).   It would also be interesting to investigate the problem of double-counting 
the interaction in the hybrid approach by carrying out a pure HF calculation,  working directly with the physical electrons rather than with KS fermions. However it is expected that the pure HF calculation will not by itself be a sufficiently accurate foundation 
for free energy calculations.  Ultimately it would be desirable to replace
hybrid DFT / Anderson impurity model approach developed here with a unified first
principles method that can simultaneously and accurately describe the
physics and chemistry of relativity, solvation, and strong electronic correlations.   

\begin{acknowledgments}
We are grateful to J. Bradley, D. L. Cox, J. Doll, E. Kim, J. Li, R. L. Martin, M. Norman,  Q. Yin and S.-C. Ying for helpful discussions.   
The work is supported in part by NSF grant DMR-0605619.
\end{acknowledgments}


%

\end{document}